\documentclass[preprint,aps,nofootinbib]{revtex4}
\usepackage{amsfonts,amsmath,amssymb,amsthm}
\usepackage{latexsym}
\usepackage{bbm,bm}
\usepackage{graphicx}

\newcommand{\ket}[1]{\lvert #1 \rangle}
\newcommand{\bra}[1]{\langle #1 \lvert}
\newcommand{\beq}{\begin{equation}}
\newcommand{\eeq}{\end{equation}}
\newcommand{\beqs}{\begin{eqnarray}}
\newcommand{\eeqs}{\end{eqnarray}}

\begin{document}

\title{Relative Entropy of Entanglement for Two-Qubit State with $z$-directional Bloch Vectors}

\author{DaeKil Park$^{1,2}$}
\affiliation{$^1$ Department of Physics, Kyungnam University, Masan, 631-701, Korea    \\
            $^2$ Department of Electronic Engineering, Kyungnam University, Masan, 631-701, Korea }

\begin{abstract}
So far there is no closed formula for relative entropy of entanglement of arbitrary two-qubit states. In this paper
we present a method, which guarantees the derivation of the relative entropy of entanglement for most states that 
have $z$-directional Bloch vectors. It is shown that the closest separable states for those states also have
$z$-directional Bloch vectors though there are few exceptions. 
\end{abstract}

\maketitle
Research into entanglement of quantum states has long history from the very beginning of quantum mechanics\cite{epr-35,schrodinger-35}. At 
that time the main motivation for the study of entanglement was to explore the non-local property of quantum mechanics. Still, this issue is not 
completely understood. Recent study on the entanglement is mainly due to its role as
a physical resource for the various quantum information processing such as teleportation\cite{bennett93}, 
quantum cryptography\cite{ekert91}, and speed-up of quantum computer\cite{vidal03-1}.

In order to quantify how much a given quantum state is entangled many entanglement measures were invented for last two decades.
Among them the most important measure seems to be the distillable entanglement\cite{bennett96} because it measures how 
a given quantum state is useful in the real quantum information processing with overcoming the effect of the noises via the 
purification protocol. In spite of its importance the analytically derivational technique for it even in the relatively
simple quantum system is not known. In fact, in order to compute the distillable entanglement we should find an optimal purification protocol. 
However, it is nontrivial problem to find the optimal protocol except very rare cases. In this reason many people tried to find more 
analytically tractable entanglement measures which may be able to provide an information on the tight upper bound of the distillable entanglement. The 
representatives constructed in this reason are entanglement of formation (EOF) \cite{bennett96} and relative entropy of 
entanglement (REE)\cite{vedral-97-1,vedral-97-2}. 

About a decade ago Wootters\cite{woot-98} found how to compute the EOF for arbitrary two-qubit states. Although still we do not
have closed formula of EOF for higher-dimensional quantum system, the Wootters' result has great impact in the study of entanglement.
One of the example for an application of the Wootters' result is to examine the role of the quantum entanglement in the complex
quantum system such as bio-system\cite{cai-08}. 
Another direction of application is to use the Wootters' result to find a truly multipartite entanglement measure. 
In this way, the three-tangle, measure for the genuine tripartite entanglement, was invented in Ref.\cite{ckw-99}. 

On the contrary, still we do not have closed formula of the REE even for the two-qubit states\cite{open05}. In order to 
understand the distillable entanglement more profoundly, therefore, it is worthwhile to investigate
the properties of the REE for the various two-qubit states. In this paper we would like to examine the REE for the states, which have 
$z$-directional Bloch vectors. We present three theorems in the following, which guarantees that the REE for the most such states can be 
computed analytically or, at least, numerically.  

The REE for state $\rho$ is defined as 
\begin{equation}
\label{ree-1}
E_R (\rho) = \min_{\sigma \in {\cal D}} S (\rho || \sigma) 
           = \min_{\sigma \in {\cal D}} \mbox{tr} \left[ \rho \ln \rho - \rho \ln \sigma \right],
\end{equation}
where ${\cal D}$ is a set of positive partial transpose (PPT) states. 
For various properties of the REE see Ref.\cite{r1,r2,r3,r4}.
If our concern is restricted into the 
two-qubit state, it is possible to regard ${\cal D}$ as a set of the separable states, because there is no
bound entangled state in the two-qubit Hilbert space. The separable state $\sigma$ in Eq.(\ref{ree-1}) is called 
the closest separable state (CSS) of $\rho$. In order for the separable state $\sigma$ to be CSS of some entangled 
states it should be edge state in the set ${\cal D}$, which means that the smallest eigenvalue of $\sigma^{\Gamma}$ 
is zero, where the superscript $\Gamma$ denotes partial transposition\footnote{The converse of this statement, 
i.e. {\it if $\sigma$ is an edge state in ${\cal D}$, there exist entangled states whose CSS are $\sigma$,} 
is not generally true}.

Although the definition of the REE is comparatively simple, the analytic computation of it is highly 
difficult problem even for the most simple two-qubit case (see chapter $8$ of Ref.\cite{open05}). Since the REE can be 
straightforwardly computed provided that the CSS is derived, this means that finding a CSS of the given entangled state 
is very difficult. Recently, however, the authors in Ref.\cite{miran-08-1} analyzed the converse procedure. When the 
edge separable state $\pi$ is full-rank, they have presented a method for deriving the entangled state $\rho$, 
whose CSS is $\pi$. Still, however, finding a CSS for the arbitrary entangled state $\rho$ is an unsolved problem.

In order to explore the issue for finding CSS or REE authors in Ref.\cite{hungsoo10-1} approached the problem from the 
geometrical point of view. To explain the main results of Ref.\cite{hungsoo10-1} briefly it is convenient to express the 
given entangled state $\rho$ in a form
\begin{equation}
\label{state-1}
\rho = \frac{1}{4} \left[ I \otimes I + {\bm r} \cdot {\bm \sigma} \otimes I + I \otimes {\bm s} \cdot {\bm \sigma}
                           + \sum_{i,j=1}^3 g_{ij} \sigma_i \otimes \sigma_j \right]
\end{equation}
where ${\bm \sigma}$ is usual Pauli matrices. The vectors ${\bm r}$ and ${\bm s}$ are Bloch vectors for each qubit and 
the tensor $g_{ij}$ represents a correlation between qubits. Since appropriate local-unitary (LU) transformation for 
each qubit can make the correlation tensor $g_{ij}$ to be diagonal, it is more convenient, without loss of generality, 
to express $\rho$ as 
\begin{equation}
\label{state-2}
\rho = \frac{1}{4} \left[ I \otimes I + {\bm r} \cdot {\bm \sigma} \otimes I + I \otimes {\bm s} \cdot {\bm \sigma}
                           + \sum_{n=1}^3 g_{n} \sigma_n \otimes \sigma_n \right].
\end{equation}					   
For example, for the four Bell states
\begin{eqnarray}
\label{bell-1}
& &\ket{\beta_1} = \frac{1}{\sqrt{2}} \left( \ket{00} + \ket{11} \right)   \hspace{1.0cm}
   \ket{\beta_2} = \frac{1}{\sqrt{2}} \left( \ket{00} - \ket{11} \right)                    \\   \nonumber
& &\ket{\beta_3} = \frac{1}{\sqrt{2}} \left( \ket{01} + \ket{10} \right)   \hspace{1.0cm}
   \ket{\beta_4} = \frac{1}{\sqrt{2}} \left( \ket{01} - \ket{10} \right),
\end{eqnarray}
the Bloch vectors ${\bm r}$ and ${\bm s}$ are vanishing and the corresponding correlation vectors become
\begin{equation}
\label{bell-2}
{\bm g}_1 = (1,-1, 1)  \hspace{1.0cm} {\bm g}_2 = (-1, 1, 1)  \hspace{1.0cm} {\bm g}_3 = (1, 1, -1)  \hspace{1.0cm} 
{\bm g}_4 = (-1,-1,-1).
\end{equation}
In Ref.\cite{hungsoo10-1} it was shown that if $\rho$ is one of Bell-diagonal, generalized Vedral-Plenio (VP) and generalized
Horodecki states, its CSS is 
\begin{equation}
\label{state-3}
\pi = \frac{1}{4} \left[ I \otimes I + {\bm r} \cdot {\bm \sigma} \otimes I + I \otimes {\bm s} \cdot {\bm \sigma}
                           + \sum_{n=1}^3 \gamma_{n} \sigma_n \otimes \sigma_n \right].
\end{equation}
The correlation vector of $\pi$, ${\bm \gamma}$, can be computed from a fact that the straight line in the correlation
vector space, which connects ${\bm \gamma} = (\gamma_x, \gamma_y, \gamma_z)$ and ${\bm g} = (g_x, g_y, g_z)$ passes through
one of Eq.(\ref{bell-2}), which is the nearest one from ${\bm g}$. Since this fact with the edge state criterion uniquely determines
the correlation vector ${\bm \gamma}$ of the CSS, it is straightforward to compute the REE for the Bell-diagonal, 
generalized VP and generalized Horodecki states. For example, let us choose the Bell-diagonal, VP and Horodecki states as following:
\begin{eqnarray}
\label{vph-1}
& &\rho_B = \lambda_1 \ket{\beta_1}\bra{\beta_1} + \lambda_2 \ket{\beta_2}\bra{\beta_2} + \lambda_3 \ket{\beta_3}\bra{\beta_3} + 
      \lambda_4 \ket{\beta_4}\bra{\beta_4}    \\  \nonumber
& & \hspace{6.0cm} (\max (\lambda_1, \lambda_2, \lambda_3, \lambda_4) = \lambda_3) 
	                                                                                                     \\   \nonumber
& &\rho_{vp} = \lambda_1 \ket{\beta_3}\bra{\beta_3} + \lambda_2 \ket{01}\bra{01} + \lambda_3 \ket{10}\bra{10}
                                                                                                         \\   \nonumber
& &\rho_{H} = \lambda_1 \ket{\beta_3}\bra{\beta_3} + \lambda_2 \ket{00}\bra{00} + \lambda_3 \ket{11}\bra{11}.
\end{eqnarray}
Following Ref.\cite{hungsoo10-1} it is easy to show that the corresponding CSS for these states are
\begin{eqnarray}
\label{vph-2}
& &\pi_B = \frac{\lambda_1}{2(1 - \lambda_3)} \ket{\beta_1}\bra{\beta_1} + 
           \frac{\lambda_2}{2(1 - \lambda_3)} \ket{\beta_2}\bra{\beta_2} + \frac{1}{2} \ket{\beta_3}\bra{\beta_3} + 
		   \frac{\lambda_4}{2(1 - \lambda_3)} \ket{\beta_4}\bra{\beta_4}                                \\   \nonumber
& &\pi_{vp} = \left(\frac{\lambda_1}{2} + \lambda_2\right) \ket{01}\bra{01} + 
              \left(\frac{\lambda_1}{2} + \lambda_3\right) \ket{10}\bra{10}	                         \\   \nonumber
& &\pi_{H} = \frac{(\lambda_1 + 2 \lambda_2) (\lambda_1 + 2 \lambda_3)}{2} \ket{\beta_3}\bra{\beta_3} + 
             \frac{(\lambda_1 + 2 \lambda_2)^2}{4} \ket{00}\bra{00} + 
             \frac{(\lambda_1 + 2 \lambda_3)^2}{4} \ket{11}\bra{11}	
\end{eqnarray}
and their REE become
\begin{eqnarray}
\label{vph-3}
& &E_r (\rho_B) = -H (\lambda_3) + \ln 2                                 \\   \nonumber
& &E_r (\rho_{vp}) = H \left( \frac{\lambda_1}{2} + \lambda_2 \right) - H (\Lambda)  
    \hspace{1.0cm} \left( \Lambda = \frac{1}{2} \left[ 1 + \sqrt{\lambda_1^2 + (\lambda_2 - \lambda_3)^2} \right] \right)
                                                                                                           \\   \nonumber
& &E_r (\rho_H) = \lambda_1 \ln \lambda_1 + \lambda_2 \ln \lambda_2 + \lambda_3 \ln \lambda_3
                   + 2 H \left( \frac{\lambda_1}{2} + \lambda_2 \right) - \lambda_1 \ln 2
\end{eqnarray}
where $H(p) \equiv -p \ln p - (1-p) \ln (1-p)$. It is worthwhile noting that $E_r (\rho_{vp})$ and $E_r (\rho_H)$ are 
invariant under the exchange of $\lambda_2$ and $\lambda_3$. In fact, one can conjecture this symmetry from the physical 
point of view. 

In this paper we would like to examine the REE for the two qubit states, whose Bloch vectors ${\bm r}$ and ${\bm s}$ are 
$z$-directional. Thus, we assume ${\bm r} = (0,0,r)$ and ${\bm s} = (0,0,s)$. For more simplicity we assume that the
first two components of the correlation vector ${\bm g}$ are identical, i.e. $g_x = g_y$. Then, the quantum state
$\rho$ can be written as 
\begin{eqnarray}
\label{entangled-1}		
\rho = \left(                       \begin{array}{cccc}
                             A_1  &  0  &  0  &  0         \\
                             0  &  A_2  &  D e^{i \varphi}  &  0         \\
                             0  &  D e^{-i \varphi}  &  A_3  &  0         \\
                             0  &  0  &  0  &  A_4
                                     \end{array}                \right)
\end{eqnarray}
where
\begin{eqnarray}
\label{para-1}
& &\hspace{1.5cm}A_1 = \frac{1+r+s+g_z}{4}  \hspace{1.0cm} A_2 = \frac{1+r-s-g_z}{4}                                 \\   \nonumber
& &A_3 = \frac{1-r+s-g_z}{4}  \hspace{1.0cm} A_4 = \frac{1-r-s+g_z}{4}  \hspace{1.0cm}	D=\frac{g_x}{2 \cos \varphi} \geq 0.
\end{eqnarray}							 
We also impose 
\begin{equation}
\label{impose-1}
D^2 > A_1 A_4
\end{equation}
to require that $\rho$ is an entangled state. 

Now we conjecture that the CSS of $\rho$ is of a form
\begin{eqnarray}
\label{css-1}
\pi = \left(             \begin{array}{cccc}
                       r_1  &  0  &  0  &  0               \\
					   0  &  r_2  &  y e^{i \varphi}  &  0               \\
					   0  &  y e^{-i \varphi} &  r_3  &  0               \\
					   0  &  0  &  0  &  r_4 
					      \end{array}                            \right)
\end{eqnarray}
with $y = \sqrt{r_1 r_4} \leq \sqrt{r_2 r_3}$. In the following we will show that most entangled states of the form (\ref{entangled-1})
have really their CSS as the form (\ref{css-1}). However, for extremely asymmetric states we will show that our conjecture is not true.

If $\pi$ is really the CSS of $\rho$, the following coupled equations should be satisfied\cite{miran-08-1}:
\begin{subequations}
\label{css-2}
\begin{equation}
\label{css-2-1}
r_1 - x \frac{r_1 r_4}{r_1 + r_4} = A_1
\end{equation}
\begin{equation}
\label{css-2-2}
r_4 - x \frac{r_1 r_4}{r_1 + r_4} = A_4
\end{equation}
\begin{equation}
\label{css-2-3}
r_2 + x \frac{2 r_1 r_4}{(r_1 + r_4) z^2 \ell} \left[ 2 r_1 r_4 \ell + (r_2 - r_3) (r_2 \ell - z) \right] = A_2
\end{equation}
\begin{equation}
\label{css-2-4}
r_3 + x \frac{2 r_1 r_4}{(r_1 + r_4) z^2 \ell} \left[ 2 r_1 r_4 \ell - (r_2 - r_3) (r_3 \ell - z) \right] = A_3
\end{equation}
\begin{equation}
\label{css-2-5}
y + x \frac{y}{(r_1 + r_4) z^2 \ell} \left[ 2 r_1 r_4 (r_2 + r_3) \ell + (r_2 - r_3)^2 z \right] = D,
\end{equation}
\end{subequations}
where $x$ is a positive parameter and
\begin{equation}
\label{css-3}
z = \sqrt{(r_2 - r_3)^2 + 4 r_1 r_4}   \hspace{1.0cm}   \ell = \ln \frac{r_2 + r_3 + z}{r_2 + r_3 - z}.
\end{equation}  
In this case one can show after tedious calculation that the REE of $\rho$ becomes
\begin{eqnarray}
\label{ree-2}
& &E_r (\rho)\equiv \mbox{tr} (\rho \ln \rho) - \mbox{tr} (\rho \ln \pi)                \\   \nonumber
& &\hspace{1.0cm} = \left( A_1 \ln A_1 + A_4 \ln A_4 + A_+ \ln A_+ + A_- \ln A_- \right)               \\   \nonumber
& & \hspace{.6cm}   - \left(A_1 \ln r_1 + A_4 \ln r_4 + \frac{A_2 + A_3}{2} \ln (r_2 r_3 - r_1 r_4) + 
     \frac{(A_2 - A_3) (r_2 - r_3) + 4 D y}{2 z \ell^{-1}} \right) 
\end{eqnarray}
where
\begin{equation}
\label{ree-3}
A_{\pm} = \frac{1}{2} \left[ (A-2 + A_3) \pm \sqrt{(A_2 - A_3)^2 + 4 D^2} \right].
\end{equation}

Now we present the following three theorems, which provide the REE and CSS of the entangled state $\rho$ 
given in Eq.(\ref{entangled-1}).

\smallskip

{\bf Theorem 1.} If $A_1 = A_4 = 0$, $E_r (\rho)$ becomes 
$$E_r (\rho) = H(A_2) - H(A_+).$$ 

\smallskip

{\bf Proof.} If $A_1 = A_4 = 0$, Eq.(\ref{css-2-1}) and Eq.(\ref{css-2-2}) give solutions 
$r_1 = r_4 = \epsilon$, where $\epsilon$ is an infinitesimal positive parameter, which will be taken to be zero
after calculation. Then, the remaining equations in Eq.(\ref{css-2}) eventually generate the following solutions:
\begin{equation}
\label{th-1-1}
r_2 = A_2  \hspace{1.0cm} r_3 = A_3 \hspace{1.0cm} x = \frac{2D}{|A_2 - A_3|} \ln \frac{\max(A_2, A_3)}{\min (A_2,A_3)}.
\end{equation}
Therefore, CSS $\pi$ in this case is 
\begin{equation}
\label{th-1-2}
\pi = A_2 \ket{01}\bra{01} + A_3 \ket{10}\bra{10}.
\end{equation}
Making use of Eq.(\ref{ree-2}) it is straightforward to compute the REE, which completes the proof.

\smallskip

As an example of theorem 1 let us consider 
\begin{equation}
\label{app-1-1}
\rho = p \ket{\psi} \bra{\psi} + q_1 \ket{01}\bra{01} + q_2 \ket{10}\bra{10}
\end{equation}
where $p + q_1 + q_2 = 1$ and $\ket{\psi} = \alpha \ket{01} + \beta \ket{10} \hspace{.3cm} (|\alpha|^2 + |\beta|^2 = 1)$.
Then the CSS of $\rho$ is 
\begin{equation}
\label{app-1-2}
\pi = (p |\alpha|^2 + q_1) \ket{01}\bra{01} + (p |\beta|^2 + q_2) \ket{10}\bra{10}
\end{equation}
and the corresponding REE is 
\begin{equation}
\label{app-1-3}
E_r (\rho) = H \left( p |\alpha|^2 + q_1 \right) - H(A_+)
\end{equation}
where
\begin{equation}
\label{app-1-4}
A_{\pm} = \frac{1}{2} \left[ 1 \pm \sqrt{p^2 + (q_1 - q_2) \left\{ 2p (|\alpha|^2 - |\beta|^2) + (q_1 - q_2) \right\}} \right].
\end{equation}
When $\alpha = \beta = 1/\sqrt{2}$, it is easy to show that Eq.(\ref{app-1-3}) reduces to the second equation of 
Eq.(\ref{vph-3}) when $\lambda_1 = p$, $\lambda_2 = q_1$ and $\lambda_3 = q_3$. 

\smallskip

{\bf Theorem 2.} If both $A_1$ and $A_4$ are not zero, and $A_2 = A_3$, the REE of $\rho$ becomes
\begin{equation}
\label{th-2-1}
E_r (\rho) = \Omega_1 - \Omega_2
\end{equation}
where 
\begin{eqnarray}
\label{th-2-2}
& &\Omega_1 = A_1 \ln A_1 + A_4 \ln A_4 + (A_2 + D) \ln (A_2 + D) + (A_2 - D) \ln (A_2 - D)           \\   \nonumber
& &\Omega_2 = A_1 \ln r_1 + A_4 \ln r_4 + A_2 \ln (r_2^2 - r_1 r_4) + D \ln \frac{r_2 + y}{r_2 - y}.
\end{eqnarray}
In Eq.(\ref{th-2-2})
\begin{eqnarray}
\label{th-2-3}
& &r_1 = \frac{1}{F} \left[2A_1 (A_1 + A_2) (A_1 + A_2 + A_4) - D^2 (A_1 - A_4) + \Delta \right]
                                                                                                        \\   \nonumber
& &r_4 = \frac{1}{F} \left[2A_4 (A_2 + A_4) (A_1 + A_2 + A_4) + D^2 (A_1 - A_4) + \Delta \right]
                                                                                                        \\   \nonumber
& &r_2 = \frac{1}{F} \left[ 2(A_1 + A_2) (A_2 + A_4) (A_1 + A_2 + A_4) - D^2 (A_1 + 2 A_2 + A_4) - \Delta \right]
\end{eqnarray}
where $y = \sqrt{r_1 r_4}$ and 
\begin{eqnarray}
\label{th-2-4}
& & F = 2 (A_1 + A_2 + A_4 + D) (A_1 + A_2 + A_4 - D)                                                  \\  \nonumber
& & \Delta = D \sqrt{D^2 (A_1 - A_4)^2 + 4 A_1 A_4 (A_1 + A_2) (A_2 + A_4)}.
\end{eqnarray}

\smallskip

{\bf Remark:} Under $A_1 \leftrightarrow A_4$, $r_2$ is invariant and, $r_1$ and $r_4$ are changed into each other. 
This fact indicates that $E_r(\rho)$ is invariant under $A_1 \leftrightarrow A_4$. The appearance of this symmetry is 
plausible from the physical point of view.

\smallskip

{\bf Proof.} Since both $A_1$ and $A_4$ are not zero, Eq.(\ref{css-2-1}) and Eq.(\ref{css-2-2}) enable us to 
express $r_4$ and $x$ in terms of $r_1$ as follows:
\begin{equation}
\label{th-2-5}
r_4 = r_1 - (A_1 - A_4)   \hspace{1.0cm}   x = \frac{(r_1 - A_1) (r_1 + r_4)}{r_1 r_4}.
\end{equation}
Since $A_2 = A_3$, Eq.(\ref{css-2-3}) and Eq.(\ref{css-2-4}) imply $r_2 = r_3$. Then inserting $r_2 = r_3$ and 
Eq.(\ref{th-2-5}) into Eq.(\ref{css-2-3}), one can express $r_2$ in terms of $r_1$ as follows:
\begin{equation}
\label{th-2-6}
r_2 = -r_1 + (A_1 + A_2).
\end{equation}
In fact, Eq.(\ref{th-2-6}) can be derived from a normalization $r_1 + 2 r_2 + r_4 = 1$. 
Finally, we consider Eq.(\ref{css-2-5}), which 
reduces to
\begin{equation}
\label{th-2-7}
y^2 - D y + r_2 (r_1 - A_1) = 0.
\end{equation}
Thus, one can express $y$ in terms of $r_1$ as a form
\begin{equation}
\label{th-2-8}
y = \frac{1}{2} \left[ D \pm \sqrt{D^2 - 4 r_2 (r_1 - A_1)} \right].
\end{equation}
Since $y^2 = r_1 r_4$, one can compute $r_1$ from Eq.(\ref{th-2-8}), which is 
\begin{equation}
\label{th-2-9}
r_1 = \frac{1}{F} \left[ 2 A_1 (A_1 + A_2) (A_1 + A_2 + A_4) - D^2 (A_1 - A_4) \pm \Delta \right].
\end{equation}
Therefore, one can easily compute $r_2$ and $r_4$ by making use of Eq.(\ref{th-2-5}) and Eq.(\ref{th-2-6}). The 
undetermined sign can be fixed by Eq.(\ref{th-2-7}). Then, Eq.(\ref{ree-2}) completes a proof of theorem 2. 

As an example of theorem 2 let us consider
\begin{equation}
\label{th-2-10}
\rho = p_1 \ket{\beta_3}\bra{\beta_3} + p_2 \ket{\beta_4}\bra{\beta_4} + q_1 \ket{00} \bra{00} + q_2 \ket{11}\bra{11}
\end{equation}
with $p_1 + p_2 + q_1 + q_2 = 1$. Then, it is straightforward to show
\begin{eqnarray}
\label{th-2-11}
& &r_1 = \frac{2 q_1 (p_1 + p_2 + 2 q_1) (p_1 + p_2 + 2 q_1 + 2 q_2) - (p_1 - p_2)^2 (q_1 - q_2) + 4 \Delta}
              { 8 (p_1 + q_1 + q_2) (p_2 + q_1 + q_2)} 
			                                                                             \\   \nonumber
& &r_2 = \frac{(p_1 + p_2 + 2 q_1) (p_1 + p_2 + 2 q_2) (p_1 + p_2 + 2 q_1 + 2 q_2) - (p_1 - p_2)^2 - 4 \Delta}
              { 8 (p_1 + q_1 + q_2) (p_2 + q_1 + q_2)}
\end{eqnarray}
where
\begin{equation}
\label{th-2-12}
\Delta = \frac{p_1 - p_2}{4} 
\sqrt{4 q_1 q_2 (p_1 + p_2 + 2 q_1) (p_1 + p_2 + 2 q_2) + (p_1 - p_2)^2 (q_1 - q_2)^2}
\end{equation}
and $r_4$ is obtained from $r_1$ by exchanging $q_1$ and $q_2$. Then it is easy to compute the REE of $\rho$ by making use
of theorem 2. When $p_2 = 0$, it is also straightforward to show that the REE of $\rho$ reduces to third equation
of Eq.(\ref{vph-3}) if one identifies $\lambda_1 = p_1$, $\lambda_2 = q_1$ and $\lambda_3 = q_2$. 		  

\smallskip

{\bf Theorem 3.} For other cases the CSS of $\rho$ can be obtained by solving an equation
\begin{equation}
\label{th-3-1}
\frac{r_2 + r_3 + z}{r_2 + r_3 - z} = \mbox{exp} \left[ \frac{z (r_1 - A_1) (r_2 - r_3)^2}
                                              {y (D-y) z^2 - 2 r_1 r_4 (r_1 - A_1) (r_2 + r_3)} \right]
\end{equation}
where
\begin{eqnarray}
\label{th-3-2}
& &r_4 = r_1 - (A_1 - A_4)                                                                            \\   \nonumber
& &r_2 = \frac{1}{4} \left[ (4 A_1 + 3 A_2 + A_3) - 4 r_1 + \sqrt{\Gamma} \right]                    \\   \nonumber
& &r_3 = \frac{1}{4} \left[ (4 A_1 + A_2 + 3 A_3) - 4 r_1 - \sqrt{\Gamma} \right]
\end{eqnarray}
and 
\begin{equation}
\label{th-3-3}
\Gamma = 16 D \sqrt{r_1 r_4} - 8 (2 A_1 + A_2 + A_3 + 2 A_4) r_1 + \left[ (A_2 - A_3)^2 + 8 A_1 (2 A_1 + A_2 + A_3) \right].
\end{equation}

{\bf Remark 1:} If Eq.(\ref{th-3-2}) and Eq.(\ref{th-3-3}) are used, one can make the lhs and rhs of Eq.(\ref{th-3-1}) in 
terms of $r_1$ only. Thus, Eq.(\ref{th-3-1}) is an equation with only one variable, which can be solved analytically or 
numerically. 

{\bf Remark 2:} If Eq.(\ref{th-3-2}) does not provide a solution for some entangled state $\rho$, this fact indicates that 
the CSS of $\rho$ is not of the form (\ref{css-1}). In this case, therefore, CSS of $\rho$ seems to have different 
structure from $\rho$.  

\smallskip

{\bf Proof.}     From Eq.(\ref{css-2-1}) and Eq.(\ref{css-2-2}) one can express $r_4$ and $x$ in terms of $r_1$, which
is exactly the same with Eq.(\ref{th-2-5}). The remaining equations in Eq.(\ref{css-2}) reduce to 
\begin{subequations}
\label{th-3-4}
\begin{equation}
\label{th-3-4-1}
2 z (r_1 - A_1) (r_2 - r_3) = \ell \left[ (r_2 - A_2) z^2 + 2 (r_1 - A_1) \left\{r_2 (r_2 - r_3) + 2 r_1 r_4 \right\} \right]
\end{equation}
\begin{equation}
\label{th-3-4-2}
2 z (r_1 - A_1) (r_2 - r_3) = \ell \left[ (A_3 - r_3) z^2 + 2 (r_1 - A_1) \left\{r_3 (r_2 - r_3) - 2 r_1 r_4 \right\} \right]
\end{equation}
\begin{equation}
\label{th-3-4-3}
2 z (r_1 - A_1) (r_2 - r_3) = \ell \frac{2 y (D-y) z^2 - 4 r_1 r_4 (r_1 - A_1) (r_2 + r_3)}{r_2 - r_3}.
\end{equation}
\end{subequations}
Since the lhs of Eq.(\ref{th-3-4}) are all identical, the rhs of them should be equal. By equalizing the rhs of 
Eq.(\ref{th-3-4-1}) with the rhs of Eq.(\ref{th-3-4-3}) one can derive
\begin{equation}
\label{th-3-5}
(r_2 - r_3) (A_2 - r_2) + 2 y (D-y) - 2 r_2 (r_1 - A_1) = 0.
\end{equation}
Similarly, one can derive
\begin{equation}
\label{th-3-6}
(r_2 - r_3) (A_3 - r_3) - 2 y (D-y) + 2 r_3 (r_1 - A_1) = 0
\end{equation}
from Eq.(\ref{th-3-4-2}) and Eq.(\ref{th-3-4-3}). 
Adding Eq.(\ref{th-3-5}) and Eq.(\ref{th-3-6}) one can express $r_2 + r_3$ in terms of $r_1$ in a form
\begin{equation}
\label{th-3-7}
r_2 + r_3 = 1 - r_1 - r_4.
\end{equation}
In fact, Eq.(\ref{th-3-7}) is a normalization for the CSS $\pi$. Combining Eq.(\ref{th-3-6}) and Eq.(\ref{th-3-7}) one can 
make the following second degree equation
\begin{equation}
\label{th-3-8}
2 r_3^2 + \left[4 r_1 - (4 A_1 + A_2 + 3 A_3)\right] r_3 + 
\left[ A_3 \left\{ (2 A_1 + A_2 + A_3) - 2 r_1 \right\} - 2 y (D-y) \right] = 0,
\end{equation}
which has roots
\begin{equation}
\label{th-3-9}
r_3 = \frac{1}{4} \left[ (4 A_1 + A_2 + 3 A_3) - 4 r_1 \pm \sqrt{\Gamma} \right].
\end{equation}
Inserting Eq.(\ref{th-3-9}) into Eq.(\ref{th-3-7}) one can express $r_2$ in terms of $r_1$ as a form
\begin{equation}
\label{th-3-10}
r_2 = \frac{1}{4} \left[ (4 A_1 + 3 A_2 + A_3) - 4 r_1 \mp \sqrt{\Gamma} \right].
\end{equation}
The undetermined sign in Eq.(\ref{th-3-9}) and Eq.(\ref{th-3-10}) can be fixed by Eq.(\ref{th-3-5}). Finally, the parameter
$r_1$ is determined by Eq.(\ref{th-3-4-3}), which reduces to Eq.(\ref{th-3-1}). This completes a proof.

\smallskip

As an example of theorem 3 let us re-consider the model which was considered by Rains in Ref.\cite{rains-99-2}, where the entangled
state is 
\begin{eqnarray}
\label{rains-1}
\rho = \left(         \begin{array}{cccc}
             \frac{1}{12}  &  0  &  0  &  0                                                                          \\
			 0  &  \frac{45907}{90000} - \frac{7 \xi}{150}  &  \frac{1201}{3750} + \frac{49 \xi}{3600}  &  0       \\
			 0  &  \frac{1201}{3750} + \frac{49 \xi}{3600}   &  \frac{29093}{90000} + \frac{7 \xi}{150}  &  0      \\
			 0  &  0  &  0  &  \frac{1}{12}
			           \end{array}                                        \right)
\end{eqnarray}
with $\xi = 1 / \ln (\frac{73}{23})$. Then Eq.(\ref{th-3-1}) directly gives $r_1 = 1/6$ and the resulting CSS of $\rho$ is 
\begin{eqnarray}
\label{rains-2}
\pi = \left(               \begin{array}{cccc}
                \frac{1}{6}  &  0  &  0  &  0                                                                        \\
				0  &  \frac{55}{144}  &  \frac{1}{6}  &  0                                                          \\
				0  &  \frac{1}{6}  &  \frac{41}{144}  &  0                                                          \\
				0  &  0  &  0  &  \frac{1}{6}
				            \end{array}                                              \right).
\end{eqnarray}
This is in agreement with Rains' result. 

As a second example of theorem 3 let us consider
\begin{equation}
\label{final-1}
\rho = p \ket{\beta_3}\bra{\beta_3} + q_1\ket{01}\bra{01} + q_2\ket{10}\bra{10} + q_3\ket{00}\bra{00} + q_4\ket{11}\bra{11}
\end{equation}
with $p=0.66$, $q_1 = 0.16$, $q_2=0.03$, $q_3=0.06$ and $q_4=0.09$. Then, Eq.(\ref{th-3-1}) cannot be solved analytically.
The numerical calculation shows that the CSS is 
\begin{equation}
\label{final-2}
\pi = p' \ket{\beta_3}\bra{\beta_3} + q_1'\ket{01}\bra{01} + q_2'\ket{10}\bra{10} + q_3'\ket{00}|\bra{00} + q_4'\ket{11}\bra{11}
\end{equation}
where $p'=0.306933$, $q_1'=0.252429$, $q_2'=0.132241$, $q_3'=0.139198$ and $q_4'=0.169198$. 

Numerical calculation shows that most entangled states of the form (\ref{entangled-1}) have their CSS as a form of (\ref{css-1}). 
However, there are states whose CSS are not of the form (\ref{css-1}). For example, the state (\ref{final-1}) with $p=0.66$, 
$q_1=0.05$, $q_2=0.07$, $q_3=0.04$ and $q_4 = 0.18$ does not have CSS of the form (\ref{css-1}). It seems to be interesting to 
derive a criterion that clarifies which entangled states $\rho$ do not have CSS of the form (\ref{css-1}). 

We have assumed {\it ab initio} that the Bloch vectors of $\rho$ are $z$-directional. In addition, we have assumed that the 
first two components of the correlation vector are equal. These assumption are chosen only for simplicity. In the near future we would 
like to re-visit the REE problem for two-qubit states with removing these assumptions as much as possible. This may shed light on 
the explicit derivation for the closed formula of REE in the two-qubit system.

\begin{acknowledgments}
Acknowledgement: This work was supported by the Kyungnam University Foundation
Grant, 2009.

\end{acknowledgments}

\end{document}